\documentclass[reprint,superscriptaddress,twocolumn,nofootinbib,notitlepage,10pt]{revtex4}
\usepackage[utf8]{inputenc}

\usepackage{amsmath,amssymb}

\usepackage{nameref,hyperref}

\makeatletter
\renewcommand{\@biblabel}[1]{\quad#1.}
\makeatother

\date{}

\usepackage{lastpage,fancyhdr,graphicx}
\usepackage{epstopdf}
\usepackage{color}
\begin{document}
%\vspace*{0.35in}

% Please capitalize all terms in the title except conjunctions, prepositions, and articles.
%\begin{flushleft}
%{\Large
\title{Bounds on transient instability for complex ecosystems}
%}
%\newline
\author{Francesco Caravelli}
\affiliation{Invenia Labs, 27 Parkside Place, Cambridge, CB1 1HQ, UK}
\affiliation{London Institute of Mathematical Sciences, 35a South Street, London, W1K 2XF, UK}
\affiliation{ Department of Computer Science, University College London, Gower Street, London WC1E 6BT, UK}

\author{Phillip Staniczenko}
\affiliation{Department of Biology, University of Maryland, College Park, Maryland, MD 20742, USA}
\affiliation{National Socio-Environmental Synthesis Center (SESYNC), Annapolis, MD 21401, USA}
\begin{abstract}
Stability is a desirable property of complex ecosystems.
If a community of interacting species is at a stable equilibrium point then it is able to withstand small perturbations
to component species' abundances without suffering adverse effects.
In ecology, the Jacobian matrix evaluated at an equilibrium point is known as the community matrix,
which describes the population dynamics of interacting species.
A system's asymptotic short- and long-term behaviour can be determined from eigenvalues derived from the community matrix.
Here we use results from the theory of pseudospectra to describe intermediate, transient dynamics.
We first recover the established result that the transition from stable to unstable dynamics includes a region of `transient instability',
where the effect of a small perturbation to species' abundances---to the population vector---is amplified before ultimately decaying.
Then we show that the shift from stability to transient instability can be affected by uncertainty in, or small changes to,
entries in the community matrix,
and determine lower and upper bounds to the maximum amplitude of perturbations to the population vector.
Of five different types of community matrix, we find that amplification is least severe when predator-prey interactions dominate.
This analysis is relevant to other systems whose dynamics can be expressed in terms of the Jacobian matrix.
Our results will lead to improved understanding of how 
multiple perturbations to a complex system may irrecoverably break stability.
\end{abstract}

% Insert additional author notes using the symbols described below. Insert symbol callouts after author names as necessary.
% 
% Remove or comment out the author notes below if they aren't used.
%

% Additional Equal Contribution Note
% Also use this double-dagger symbol for special authorship notes, such as senior authorship.
%\ddag These authors also contributed equally to this work.

% Use the asterisk to denote corresponding authorship and provide email address in note below.
%*francesco.caravelli@invenialabs.co.uk

%\end{flushleft}
\maketitle

% Please keep the Author Summary between 150 and 200 words
% Use first person. PLOS ONE authors please skip this step. 
% Author Summary not valid for PLOS ONE submissions.   
%\section*{Author Summary}
%If an ecosystem is perturbed, for example, through the loss of a small number of individuals of a component species, 
%then one of two outcomes is typically assumed: 
%either the perturbation quickly dissipates and the ecosystem returns to its original state---the ecosystem is stable; 
%or the perturbation causes the ecosystem to move to a new state, possibly with fewer individuals or species---the ecosystem is unstable. 
%Here we characterise a third possibility: transient instability. In this case, a perturbation is first amplified before eventually dissipating. 
%Transient instability is particularly relevant during times of rapid environmental change. 
%This is because additional perturbations during the amplification period may prevent 
%the ecosystem from returning to its original state, along with the unexpected loss of species and biodiversity.

%\linenumbers
\section{Introduction}
From the perspective of local stability analysis, if an ecosystem
is close to a stable equilibrium point then the effect of a small perturbation, such as the  loss of individuals from a population,
will eventually decay and the system will return to its original equilibrium point~\cite{PimmNature1984,MontoyaNature2006}.
But if the ecosystem is at an unstable equilibrium point then the perturbation will lead to the system settling at a new equilibrium point,
possibly with fewer individuals or even species~\cite{MayNature1977,McNaughtonNature1978}.
In theory, ecosystems with large numbers of species and interactions are more difficult to stabilise~\cite{MayNature1972}.
However, many ecosystems contain vast biodiversity~\cite{YodzisNature1981,McCannNature2000}.
Reconciling this finding with local stability analysis has motivated ecologists for over 40 years~\cite{AllesinaTangPopEco2015}.

Recently, stability criteria were extended from randomly assembled communities to include those with
more realistic compositions of mutualistic, competitive and predator-prey interactions~\cite{AllesinaTangNature2012}.
These criteria showed that communities in which predator-prey interactions dominate are more likely to be stable.
It was then shown, using empirical food webs, that the distribution and correlation of interaction strengths has
a greater effect on stability than topology: how species interact with one another is more important 
than who they interact with~\cite{NeutelThorneEcolLett2014,TangEcolLett2014}.

Stability is a long-term concept: it indicates whether a system will, at some point in the future,
return to the same state as before a perturbation~\cite{NeubertCaswellEcology1997}.
Reactivity, on the other hand, indicates how a system will respond immediately 
after a perturbation has been applied~\cite{NelsonShnerbPRE1998,CaswellNeubertJDiff2005,VerdyCaswellBullMatBiol2008,NeubertEcology2009,SnyderTheorPopBiol2010}.
A stable system can be non-reactive, meaning that a perturbation to species' abundances dies down immediately, or reactive, meaning that a perturbation is first amplified before eventually decaying
(whether a particular perturbation is amplified in practice depends on which species
are perturbed and by how much~\cite{NelsonShnerbPRE1998}).
Reactivity criteria for large ecosystems show that communities on the verge of instability exhibit
reactive dynamics~\cite{TangAllesinaFEE2014}, and  
identifying a system as reactive has been proposed as an early-warning signal for 
population collapse~\cite{SchefferNature2009,SchefferScience2012,VeraartNature2012,DaiScience2012,DaiNature2013}.

The starting point for deriving criteria for both stability and reactivity is the community matrix~\cite{LevinsBook1968}.
%This is the Jacobian matrix representing the effect of interspecific interactions on population density evaluated
%at an equilibrium point.
A spectral decomposition of the community matrix provides information on the asymptotic behaviour of the system
for stability ($t \to \infty$) and reactivity ($t \to 0$).
But so far, little information has been extracted from the community matrix regarding transient dynamics:
how the system evolves after a perturbation and before it either returns to equilibrium or 
becomes unstable~\cite{ChenCohenPRSB2001,NeubertEcolMod2004,HastingsTREE2004}.

Reactive dynamics are not possible if the community matrix ${\bf M}$ is normal, i.e., ${\bf M} {\bf M}^\dagger={\bf M}^\dagger {\bf M}$, where
${\bf M}^\dagger$ is the adjoint of ${\bf M}$~\cite{TrefethenSIAMREV1997}. 
But if ${\bf M}$ is a non-normal matrix, as is usually the case in analyses of realistic ecosystems, then transient dynamics may substantially differ from the asymptotic behaviour suggested by 
the eigenvalues of ${\bf M}$. 
In addition, small changes to the entries of non-normal ${\bf M}$ can cause an otherwise stable matrix to become unstable~\cite{TrefethenSIAMREV1997}.
In such cases,
the dynamics implied by non-normal matrices are better described by pseudospectra,
which detail the neighbourhood of eigenvalues in the complex plane
for different average changes to the entries in ${\bf M}$~\cite{TrefethenEmbreeBook2005}.

Here we formalise the transition from stability to instability in terms of pseudospectra. 
Using this approach, we consider the effect on dynamics of two kinds of perturbation:
more commonly studied perturbations to the equilibrium abundance of species (to the population vector) and less commonly
studied perturbations to the entries in ${\bf M}$ (which could be interpreted as uncertainty in, or small changes to, species' interaction strengths~\cite{BarabasAllesinaJRSI2015}).
We describe critical values for community properties separating three regimes:
stable and non-reactive dynamics,
stable and reactive dynamics---`transient instability'---and unstable dynamics.
We show that system dynamics at the boundary between non-reactive stability and transient instability
can be affected by perturbations to entries of the community matrix. 
And, given a perturbation to the equilibrium abundance of species, we provide upper and lower bounds to the maximum amplification of such perturbations during transient instability.
This allows us to sketch out the transient dynamics of complex ecosystems using only information
from the community matrix.
Finally, we compare the properties of community matrices representing ecological communities with five different types of interaction structure: random, mutualism, competition, mixture of mutualism and competition, and predator-prey.

\section{Methods}
\subsection{Local stability analysis} Here we consider an ecological community of $S$ species 
for which their population densities at time $t$ are given by the vector ${\bf Y}(t)$,
as in Tang \& Allesina~\cite{TangAllesinaFEE2014}.
The dynamics of the population vector ${\bf Y}$ can be described by a system of coupled differential equations
\begin{equation}
\frac{d {\bf Y}}{dt}={\bf f}({\bf Y})
\end{equation}
where ${\bf f}=[f_1, f_2 \cdots, f_S]^T$ is a vector of linear or nonlinear functions. 
An ecologically-relevant equilibrium point is a non-negative vector {\bf Y*} such that
\begin{equation}
{\bf f}({\bf Y^*})={\bf 0}
\end{equation}
The community matrix ${\bf M}$ is defined as
\begin{equation}
M_{ij}=\frac{\partial f_i}{\partial Y_j}\bigg|_{{\bf Y}={\bf Y^*}}
\end{equation}
which is the Jacobian matrix evaluated at an equilibrium point~\cite{LevinsBook1968}. 
It is well known that an equilibrium point is (locally and asymptotically) stable if any infinitesimally small deviation, $\Delta {\bf Y}(0)$,
eventually decays to zero, i.e.,  $\lim_{t\rightarrow \infty} \Delta {\bf Y}(t)=0$~\cite{LevinsBook1968}.
In the vicinity of an equilibrium point, the time evolution of a perturbation can be described by
\begin{equation}
\Delta {\bf Y}(t) = e^{{\bf M}t}\Delta {\bf Y}(0)
\label{timeevo}
\end{equation}
Therefore, the spectrum of the community matrix ${\bf M}$ is clearly relevant for determining local stability.
%The criterion for having a local equilibrium point, 
If $\Lambda(${\bf M}$)$ is the set of eigenvalues of ${\bf M}$, 
then an equilibrium point is stable if all eigenvalues have negative real part, i.e., $\textit{Re}(\lambda)<0 \; \forall\ \lambda \in \Lambda(\bf M)$~ \cite{MayNature1972,AllesinaTangNature2012}.

\subsection{Generative models for community matrices}
\label{Sec:GenModels}
We parameterise community matrices using four quantities:
$S$, $C$, $\mu$ and $\sigma$;
where $S$, as above, is the number of species, $C$ is the connectance (the fraction of 
realised interactions among species), $\mu$ is the strength of intraspecific interactions
and $\sigma$ is the standard deviation of the strength of interspecific interactions~\cite{AllesinaTangNature2012}.
We assume that populations are self-regulating and so
$M_{ii}=-\mu$, where $\mu>0$.
Non-normal community matrices with different types of interaction---representing different types
of ecological community---are generated by
sampling off-diagonal entries ($M_{ij}$, interspecific interactions) from different bivariate distributions.
Having specified a particular distribution, 
stability criteria can be expressed in terms of $S$, $C$, $\mu$ and $\sigma$.
Based on these criteria, it has been shown that predator-prey community matrices are the most stable,
followed by random, competition, mixture and mutualism~\cite{AllesinaTangNature2012}.
Generative models for these community matrices are described below.

\textit{Random.} Each off-diagonal entry is sampled independently from a normal distribution $\mathcal{N}(0,\sigma)$
with probability $C$, and otherwise $M_{ij} = 0$ with probability $1-C$.
%The probability of having an eigenvalue with positive real part and therefore instability is close to zero if $\sigma \sqrt{SC}< \mu$. 

\textit{Mutualism.} Each off-diagonal pair $(M_{ij},M_{ji})$ is sampled from a half-normal distribution $|\mathcal{N}(0,\sigma)|$
with probability $C$, and both entries are zero otherwise.
These community matrices have a $(+,+)$ sign structure for off-diagonal pairs. 
%The stability criterion is $\sigma (S-1) C< \frac{\sqrt{\pi} \mu}{\sqrt{2}}$.

\textit{Competition.} Each off-diagonal pair $(M_{ij},M_{ji})$ is sampled from a half-normal distribution $-|\mathcal{N}(0,\sigma)|$
with probability $C$, and both entries are zero otherwise.
These community matrices have a $(-,-)$ sign structure for off-diagonal pairs. 
%The stability criterion is $\sigma [\sqrt{S C} (1+\frac{2-2C}{\pi-2C})\sqrt{\frac{\pi-2C}{\pi}}+C\frac{\sqrt{2}}{\sqrt{\pi}})]<\mu$.

\textit{Mixture of mutualism and competition.} Each off-diagonal pair $(M_{ij},M_{ji})$ is sampled from a half-normal distribution $|\mathcal{N}(0,\sigma)|$
with probability $C/2$ or $-|\mathcal{N}(0,\sigma)|$ with probability $C/2$, and both entries are zero otherwise.
These community matrices have a $(+,+)$ or $(-,-)$ sign structure for off-diagonal pairs. 
%The stability criterion is $\sigma \sqrt{S C}<\mu \frac{\pi}{\pi+2}$.

\textit{Predator-prey.} The first entry in an off-diagonal pair is sampled from a half-normal distribution $|\mathcal{N}(0,\sigma)|$
and the second entry from $-|\mathcal{N}(0,\sigma)|$ with probability $C/2$, or with the half-normal distributions reversed with
probability $C/2$, and both entries are zero otherwise.
These community matrices have a $(+,-)$ or $(-,+)$ sign structure for off-diagonal pairs.
%The stability criterion is $\sigma \sqrt{S C}<\mu \frac{\pi}{\pi-2}$.

\subsection{Pseudospectra and transient instability}
%Non-normal community matrices can exhibit amplification of perturbations \textcolor{magenta}{for stable as well as unstable equilibria}.
In general, the eigenvalues of ${\bf M}$ satisfy the following definition:
\begin{equation}
\Lambda({\bf M}) = \{z \in \mathbb{C} : \text{det}(z {\bf I}-{\bf M}) = 0\}
\end{equation}
meaning that if $z$ is an eigenvalue of ${\bf M}$ then by convention the
norm of $(z {\bf I}-{\bf M})^{-1}$ is defined to be infinity (see Chapter~$I.1$ in~\cite{TrefethenEmbreeBook2005}).
But if $||(z {\bf I}-{\bf M})^{-1}||$ is finite and very large, as is often the case with perturbed non-normal matrices, 
then the pseudospectrum of ${\bf M}$ must be considered.
The `$\epsilon$-pseudospectrum' has several equivalent definitions that describe the eigenvalues of a matrix 
whose entries have been subject to noise of magnitude $\epsilon$ (in the sense of the matrix norm)~\cite{TrefethenSIAMREV1997}.
We use the following definition:
\begin{equation}
\Lambda_{\epsilon}({\bf M})=\{z \in \mathbb{C} : ||(z {\bf I}-{\bf M})^{-1}|| \geq \epsilon^{-1}\}
\end{equation}
If a matrix is normal then its $\epsilon$-pseudospectrum (henceforth just `pseudospectrum') 
consists of closed balls of radius $\epsilon$ surrounding the original eigenvalues of ${\bf M}$ (see Theorem~2.2 in~\cite{TrefethenEmbreeBook2005}).
As mentioned earlier, normal matrices cannot exhibit reactive dynamics:
perturbations of the population vector for a stable system decay immediately and with 
exponential profile as the system returns to its original equilibrium point.
But with non-normal matrices, pseudospectra can be much larger and more intricate and reactive dynamics are possible: perturbations of the population vector for a stable system first increase in magnitude and reach a maximum amplitude before eventually decaying (Fig.~1). 
This behaviour motivates a description of local stability analysis for community matrices in terms of pseudospectra.

\begin{figure}
\centering
\includegraphics[scale=0.27]{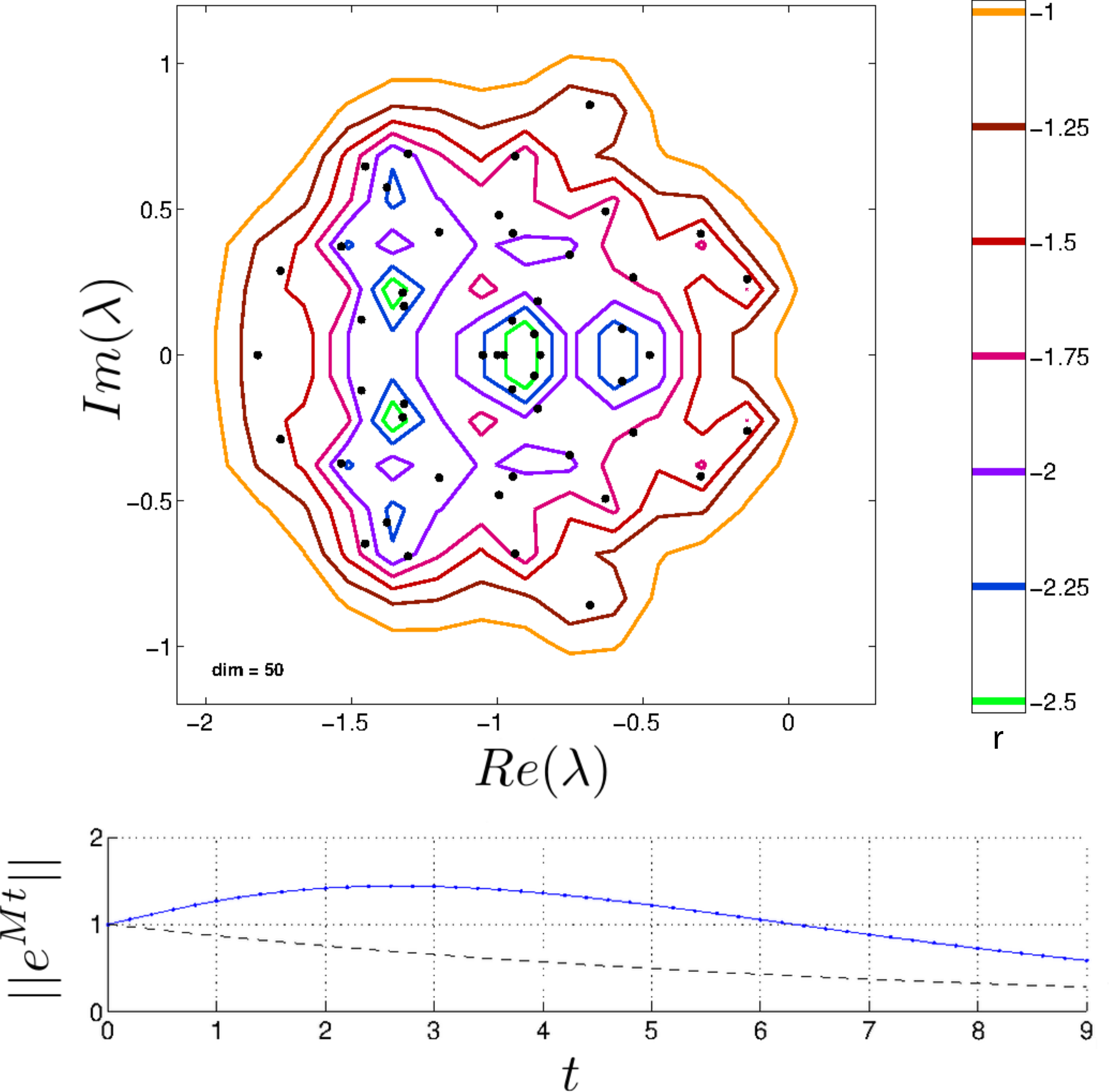}
\caption{\textit{Top:} Pseudospectrum of a random community matrix with $S=50$, $C=0.1$, $\mu=1$ and $\sigma=0.3$,
which is asymptotically stable. 
Contours in the complex plane illustrate the effect on eigenvalues of the community matrix ${\bf M}$ for noise of magnitude $\epsilon=10^{r}$~\cite{Pseudowebsite}.
The contour for $\epsilon=0.1$ crosses the imaginary axis, implying that the pseudospectral abscissa is positive and 
so transient instability is observable. 
\textit{Bottom:} Dynamics of $||e^{{\bf M} t}||$ (arbitrary units of time, see Eqn~\ref{Eqn:Bounds}). The dashed curve represents dynamics from eigenvalue analysis, whereas the solid curve represents dynamics predicted by positive $\epsilon$-pseudospectral abscissa for $\epsilon\approx 0.1$.}
\label{Fig:Pseudo}
\end{figure}

Local asymptotic stability is determined in the same way for normal and non-normal matrices. 
The `spectral abscissa' of ${\bf M}$ is defined as
\begin{equation}
\alpha({\bf M})=\sup_{z\in \Lambda({\bf M})} \text{Re}(z)
\label{Eqn:specab}
\end{equation}
where the supremum (sup) selects for the largest (real-part) of the rightmost eigenvalue in the set $\Lambda({\bf M})$.
Stability is guaranteed for $\alpha({\bf M}) < 0$.
If ${\bf M}$ is normal, then $||e^{{\bf M} t}|| = e^{\alpha({\bf M}) t}$ and dynamics are completely
described by $\alpha({\bf M})$ (see Eqn~\ref{timeevo}). 
Otherwise, the dynamics implied by ${\bf M}$ can be more complicated:
\begin{equation}
e^{\alpha({\bf M}) t} \leq ||e^{{\bf M} t}|| \leq \kappa({\bf V}) e^{\alpha({\bf M}) t}
\label{Eqn:Bounds}
\end{equation}
where the columns of matrix ${\bf V}$ are the eigenvectors of {\bf M},
and $\kappa({\bf V})=||{\bf V}||\cdot||{\bf V}^{-1}||$ is known as the conditioning of ${\bf V}$~\cite{KreissMathComp1968,LeVequeTrefethenAMM1984,LeVequeTrefethenBIT1984,TrefethenBauSIAM1997}.
The conditioning provides a bound from above---an upper bound---to the maximum amplitude 
of a perturbation of the population vector (it is worth noting that $\kappa(\bf V)$ does not provide any information 
about the time at which the perturbation reaches its maximum amplitude).

In complement to stability is reactivity,
which describes the behaviour of a system close to $t=0$, at the application of a perturbation.
The `numerical abscissa' of ${\bf M}$ is defined as
\begin{equation}
\omega({\bf M}) = \frac{d}{dt}|| e^{{\bf M} t} ||\Big|_{t=0} = \sup_{z\in \Lambda({\bf H})} \text{Re}(z)
\label{Eqn:numabscissa}
\end{equation}
where ${\bf H} = \frac{{\bf M}+{\bf M}^t}{2}$~\cite{NelsonShnerbPRE1998,CaswellNeubertJDiff2005,VerdyCaswellBullMatBiol2008,NeubertEcology2009,SnyderTheorPopBiol2010}.
The numerical abscissa is the maximum initial amplification rate following an infinitesimally small perturbation
to the population vector.
Dynamics are non-reactive if $\omega({\bf M}) < 0$
and may be reactive if $\omega({\bf M}) \ge 0$.
A stable system can be either reactive or non-reactive, but an unstable system is necessarily reactive.

With non-normal matrices, perturbations to the entries of ${\bf M}$ 
can affect whether a system is stable and non-reactive or stable and reactive.
In other words, perturbations to the entries of {\bf M} can affect how a system responds to perturbations to the population vector.
The effect of such perturbations to {\bf M} is not covered by Eqn~\ref{Eqn:numabscissa}.
However, we can study the pseudospectrum of a community matrix to better understand system dynamics between
the limits of reactivity and stability.
In what follows, we use the theory of pseudospectra to relate uncertainty in, or small changes to, the entries of ${\bf M}$
to bounds on the amplification of perturbations of the population vector.

The `$\epsilon$-pseudospectral abscissa' of ${\bf M}$ is defined as
\begin{equation}
\alpha_{\epsilon} ({\bf M})=\sup_{z \in \Lambda_\epsilon({\bf M})} \text{Re}(z)
\label{Eqn:pseudodef}
\end{equation}
which is the largest real-part eigenvalue of the pseudospectrum of ${\bf M}$ for a given amount of noise $\epsilon$.
The $\epsilon$-pseudospectral abscissa provides a lower bound to
the maximum amplification of a perturbation of the population vector (see Eqn~14.6~in~\cite{TrefethenEmbreeBook2005}):
\begin{equation}
\sup_{\epsilon\geq 0} \frac{\alpha_\epsilon({\bf M})}{\epsilon} \leq \sup_{t \geq 0} ||e^{{\bf M} t}||
\label{Eqn:lowerbound}
\end{equation}
and therefore the function
\begin{equation}
f_{\bf M}(\epsilon)=\frac{\alpha_\epsilon({\bf M})}{\epsilon}
\label{Eqn:fM}
\end{equation}
is useful for understanding transient dynamics~\cite{Afootnote}.
In the literature on pseudospectra, $\sup_{\epsilon\geq 0} f_{{\bf M}}(\epsilon)\equiv \mathcal K({\bf M})$ 
is known as the Kreiss constant~\cite{KreissMathComp1968,LeVequeTrefethenBIT1984}.
Eqns ~\ref{Eqn:pseudodef},~\ref{Eqn:lowerbound}~and~\ref{Eqn:fM}~are 
useful because they relate perturbations to the matrix norm---small changes to the elements of the community matrix
as described by the noise parameter $\epsilon$---to the effect of perturbations
to the population vector (compare Eqns~\ref{Eqn:specab}~and~\ref{Eqn:pseudodef}).
For a given community matrix,
as the size of a \emph{matrix} perturbation is increased from zero
there may be some critical value $\epsilon^*$ at which $f_{\bf M}(\epsilon^*)=1$.
In the pseudospectrum, this is illustrated by the $\epsilon^*$-contour crossing the imaginary axis (Fig.~1).
At this point, perturbations to the equilibrium population vector begin to be amplified.

For a stable and non-reactive system, perturbations to the population vector are not amplified and
the system always returns to its original equilibrium point.
For an unstable and necessarily reactive system, perturbations are amplified
and the system may move to a new equilibrium point.
But for a stable and reactive system, perturbations are first amplified
before the system eventually returns to its original equilibrium point---this is transient instability.
Now that we can compute upper (Eqn~\ref{Eqn:Bounds}) and lower bounds (Eqn~\ref{Eqn:lowerbound}) for amplifications,
we are in a position to compare the transient dynamics of different types of ecological community
described by non-normal community matrices.

\section{Results}
We generated multiple sets of community matrices with $C=0.1$, $\mu=1$ and various combinations of $S$ and $\sigma$
for the five generative models.
We first consider lower and upper bounds to the maximum amplitude of perturbations to the population vector for
random community matrices, before turning our attention to the other types of interaction.
The data required to reproduce the plots in this article are available at \cite{CaravelliStaniczenkoData}.

\subsection{Lower bound for random community matrices}
We numerically evaluated the $\epsilon$-pseudospectral abscissa using the recently proposed subspace method~\cite{KressnerVandereyckenSIAM2014}.
Consider an ensemble of community matrices generated with random interaction type and $S=100$ and $\sigma=0.3$,
which is just below the threshold for instability ($\sigma_{\text{c}} = \frac{\mu}{\sqrt{SC}} = \frac{1}{\sqrt{10}} \approx 0.31$).
We found that the average value of $f_{\bf M}(\epsilon)$ (Eqn~\ref{Eqn:fM}) monotonically increases as a
function of $\epsilon$ and eventually saturates.
At $\epsilon^* \approx0.085$ the curve crosses one, at which point 
perturbations are amplified and transient instability may be observable.
The function $f_{\bf M}(\epsilon)$ converges for all
asymptotically stable community matrices considered here.
%This is a feature shared by all the ensembles considered in this paper, and which allows to evaluate 
%$\text{sup}_\epsilon f_{\bf M}(\epsilon)$ only for large values of $\epsilon$. 
%In the case unstable systems, this is not true anymore, as for $\epsilon\approx 0$, $\alpha_\epsilon>0$, 
%and thus the maximum value of the function $f_{\bf M}(\epsilon)$ diverges; 
%the value evaluated asymptotically is only true for the case of stable systems.

In general, we identify regions of stability, transient instability and instability
by plotting $\sup_{\epsilon\geq 0} \frac{\alpha_\epsilon({\bf M})}{\epsilon}$
(Eqn~\ref{Eqn:lowerbound}; in practice, we plot $f_{\bf M}(\epsilon)$ for large values of $\epsilon$) as $\sigma$ is varied~(Fig.~2).
Similar regions can be identified as $S$ is varied while $\sigma$ is held constant (results not shown).
In the stable region, there is no perturbation to the community matrix large enough (that can still be considered infinitesimally small)
such that $\sup_{\epsilon\geq 0} \frac{\alpha_\epsilon({\bf M})}{\epsilon} > 1$,
and so perturbations are never amplified.
%\textcolor{green}{and its average over ${\bf M}$ over all the community matrices generative models considered here is always a convex function of $\epsilon$.}.
At some critical point, $\sigma_{\text{ti}}$, there is a level of matrix noise $\epsilon=\epsilon^*$
above which perturbations to the population vector are amplified before decaying.
As $\sigma$ increases in the region of transient instability,
$\epsilon^*$ decreases until it reaches zero
at $\sigma_{\text{c}}$.
At this point, system dynamics are guaranteed to be asymptotically unstable
and any infinitesimally small perturbation to the population vector is amplified
(without necessarily returning to the original equilibrium point).
In the unstable region, $f_{\bf M}(\epsilon)$ diverges
and corresponding values for the lower bound should be treated with caution.

\begin{figure}[t!]
\centering
\includegraphics[scale=0.3]{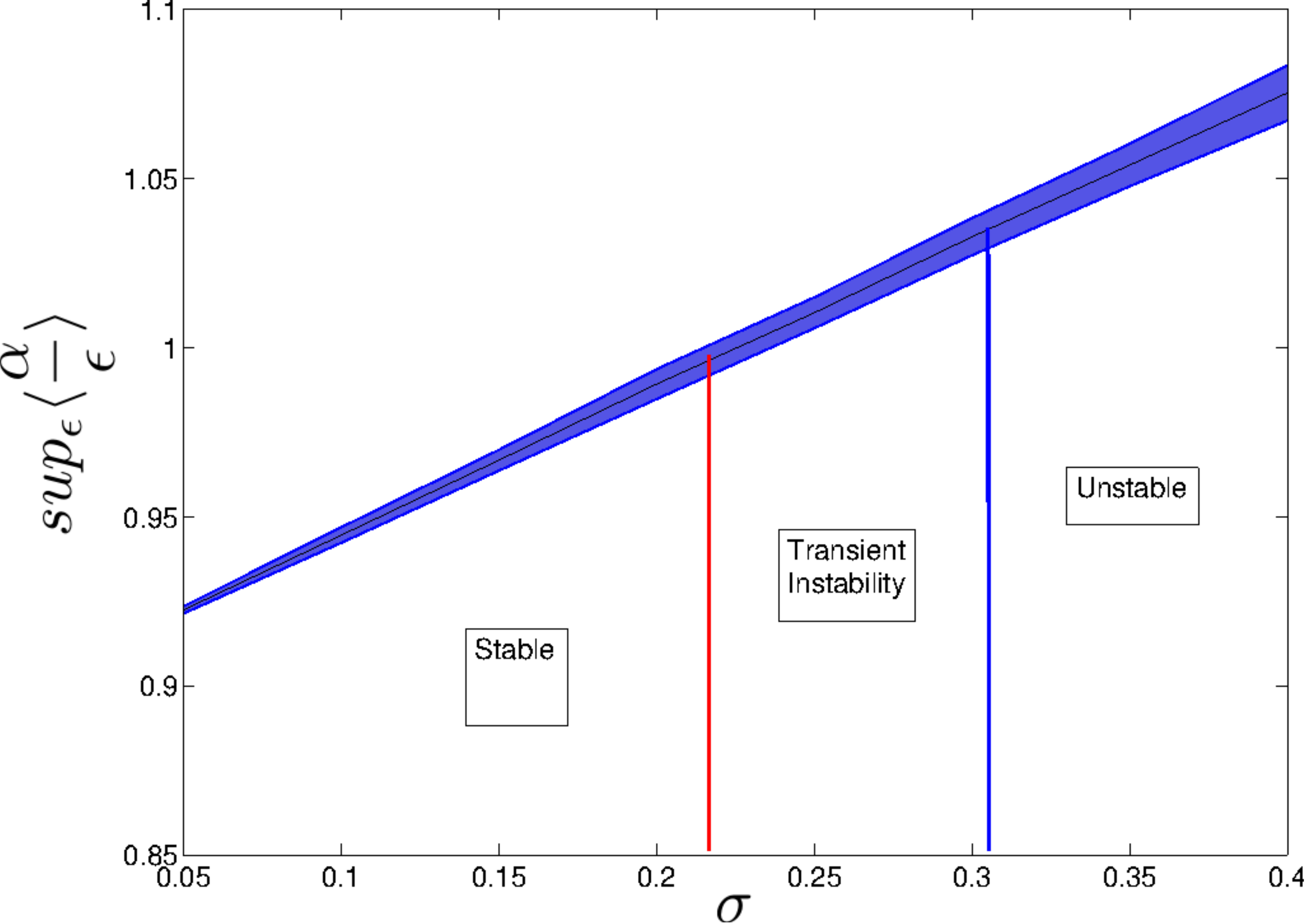}
\caption{Regions of stability, transient instability and instability for a random community matrices with 
$S=100$, $C=0.1$ and $\mu=1$ as $\sigma$ is varied.
The y-axis is the lower bound of the maximum amplitude of perturbations to the population vector~(Eqn~\ref{Eqn:lowerbound}).
Transient instability is observable as the curve crosses one at $\sigma_{\text{ti}} \approx 0.22$ and
instability is reached at $\sigma_{\text{c}} = \frac{\mu}{\sqrt{SC}} = \frac{1}{\sqrt{10}} \approx 0.31$.
At the threshold of instability, the lower bound of the maximum amplitude is $\textsc{lb}(\sigma_{\text{c}}) =1.046 \pm 0.006$ (mean~$\pm$~std).
The shaded area represents the standard error over 100 realisations.}
\label{Fig:LB}
\end{figure}

The critical point for transient instability with $S=100$ is $\sigma_{\text{ti}} \approx 0.22$. 
This is very close to the value given by reactivity criteria based on the numerical abscissa:
$\sigma_{\text{R}} = \frac{1}{\sqrt{2SC}} = \frac{1}{\sqrt{20}}$~\cite{TangAllesinaFEE2014}.
Indeed, both approaches determine whether perturbations to the population vector are amplified
based on eigenvalues related to ${\bf M}$.
As a point of difference, however, the pseudospectral approach allows for an additional treatment of uncertainty in, or small changes to, entries of the community matrix. 
For a given set of parameters, the numerical abscissa only indicates whether amplification is possible,
whereas the pseudospectrum, through the $\epsilon$-pseudospectral abscissa, 
also indicates whether amplification is possible given small changes to the strengths of interactions among species in the community.

\begin{figure}[t!]
\centering
\includegraphics[scale=0.3]{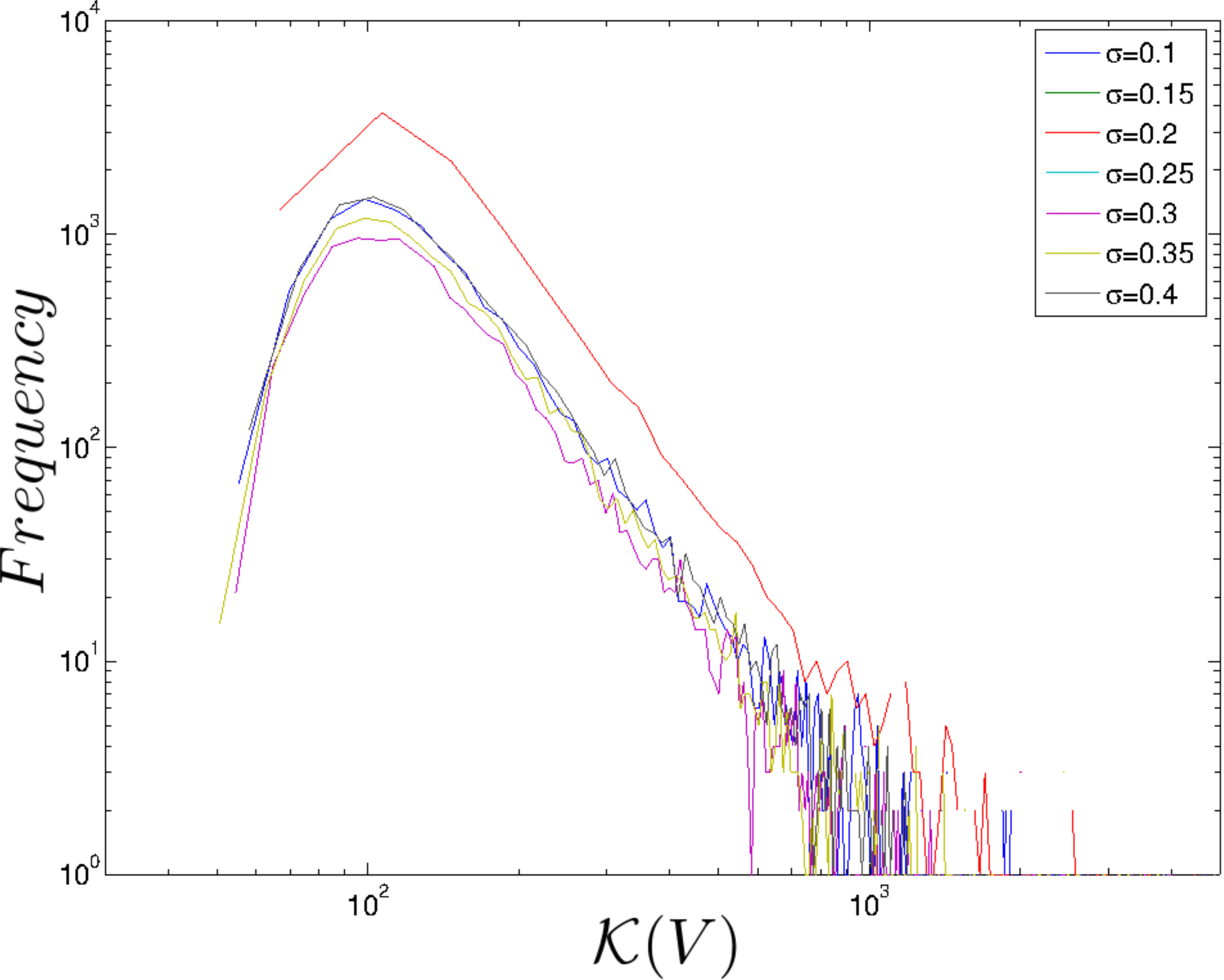}
\caption{Distribution of upper bounds of the maximum amplitude of perturbations to the population vector~(Eqn~\ref{Eqn:Bounds})
for random community matrices generated with $S=100$, $C=0.1$ and $\mu=1$ and seven values of $\sigma$ (10,000 realisations).
Distributions are fat-tailed and the slope of the tail does not change with $\sigma$.}
\label{Fig:UB}
\end{figure}

\subsection{Upper bound for random community matrices}
We plot the frequency distribution of $\kappa({\bf V})$ (Eqn~\ref{Eqn:Bounds}) 
for various combinations of $S$ and $\sigma$
to investigate the upper bound to the maximum amplitude of perturbations 
of the population vector.
In general, distributions are strongly peaked and fat-tailed~(Fig.~3).
This indicates that very large amplification is possible even for 
very small perturbations.
The location of the peak changes very little as $\sigma$ increases,
but shifts rightwards as $S$ increases (results not shown).
The slope of the tail does not change much as either $S$ or $\sigma$ is varied.
With $S=100$ and $\sigma=\sigma_{\text{c}}=0.31$, 
the peak in the distribution of upper bound values is $\textsc{ub}_{\text{peak}}(\sigma_{\text{c}})\approx95$
and the maximum value in the tail is $\textsc{ub}_{\text{tail}}(\sigma_{\text{c}})\sim1000 $.
When a power law is fit to the tail, $f(x) \propto x^{-\alpha}$,
the exponent is $\alpha \approx 2.9$. 

\begin{table}
\caption{Properties of community matrices with $S=100$, $C=0.1$, $\mu=1$.}
\label{Table:Properties}
\begin{tabular}[t]{l|ccccccc}
\multicolumn{6}{c}{}\\
Type & $\sigma_{\text{ti}}$ & $\sigma_{\text{c}}$ & $\textsc{lb}(\sigma_{\text{c}})$ & $\textsc{ub}_{\text{peak}}(\sigma_{\text{c}})$ & $\textsc{ub}_{\text{tail}}(\sigma_{\text{c}})$ & $\alpha$\\\hline
Mutualism & $0.11$ & $0.16$ & $1.02$ & $100$ & $\sim 1000$ & 3 \\
Mixture & $0.17$ & $0.19$ & $1.02$ & $77$ & $\sim 1000$ & 2.7\\
Competition & $0.17$ & $0.20$ & $1.02$ & $100$ & $\sim 1000$ & 3 \\
Random & $0.22$ & $0.31$ & $1.03$ & $95$ & $\sim1000$ & 2.9\\
Predator-prey & $0.37$ & $0.87$ & $1.10$ & $60$ & $\sim 500$ & 3.4\\
\end{tabular}
\end{table}

\subsection{Community matrices with different types of interaction}
The region of transient instability varies for different types of interaction,
as do lower and upper bounds for amplification~(Table~1).
Transient instability becomes observable with smallest $\sigma_{\text{ti}}$ with mutualism,
followed by mixture, competition, random and predator-prey.
This order is the same as for the threshold for instability, $\sigma_{\text{c}}$.
However, the size of the region of transient instability, $\sigma_{\text{c}} - \sigma_{\text{ti}}$, has a different order: 
predator-prey is largest, followed by random, mutualism, competition and mixture.
The pattern is similar if $S$ is varied while $\sigma$ is held constant (results not shown).
As expected, these findings are consistent with earlier results
based on the numerical abscissa and the correlation between off-diagonal entries in a
community matrix~\cite{TangAllesinaFEE2014}.

Predator-prey community matrices are relatively stable and exhibit the largest
range of parameter values for transient instability.
The lower bound to the maximum amplitude of perturbations of the population vector also reaches its largest value 
among the five types of interaction for predator-prey community matrices.
However, the peak in the distribution of upper bounds is at lower amplification and the slope of the tail is steeper~(Table~1).
This implies that perturbations are typically amplified less severely compared to the other types of interaction
and the very largest possible amplitudes are not as large.

Mutualism $(+,+)$ and competition $(-,-)$ have different critical points for transient instability and instability,
but similar bounds to the maximum amplitude of perturbations of the population vector.
Interestingly, the peak in the distribution of upper bounds is at lower amplification for community matrices 
with a mixture of these two interaction types.
The largest upper bound, $\textsc{ub}_{\text{tail}}(\sigma_{\text{c}})$, however, is similar to mutualism and competition,
so the exponent, $\alpha$, is shallower.

\section{Discussion}
Here we described transient instability for non-normal community matrices using local stability analysis and pseudospectra.
We showed how the shift from stable and non-reactive dynamics to transient instability
changes if perturbations are applied to the community matrix.
We also characterised how perturbations of the population vector are amplified during periods of transient instability for different types of interaction.
We found an early, sharp and severe transition between stability and instability with mutualism, mixture and competition,
but a later, longer and less severe transition with predator-prey community matrices.

In this study, we assumed a random topology of interactions between species.
Although the correlation between interaction strengths---and therefore the predominant type of interaction
in a community matrix---may be more important than topology for stability~\cite{NeutelThorneEcolLett2014,TangEcolLett2014},
it remains to be seen whether this is the case with transient instability.
Nevertheless, it is likely that the particular trajectory of a perturbed system is sensitive to
topology, and, of course, the \emph{direction} of initial perturbation of the population vector.
Understanding transient dynamics at this level of detail requires analysis of pseudoeigenvectors in addition to pseudoeigenvalues~(see Chapter~$I.4$ in~\cite{TrefethenEmbreeBook2005}).

Local stability analysis is only one approach to understanding the capacity for ecosystems to withstand external shocks~\cite{DonohueEcolLett2013,RohrScience2014}.
It will be informative to compare how the time evolution of the same shock to the same system
is assessed under different approaches to measuring the `stability', `persistence' or `resilience' of ecosystems~\cite{NeubertCaswellEcology1997}.

Stability, in principle, promises a degree of certainty that biodiversity will not be lost~\cite{PimmNature1984,MontoyaNature2006}.
Reactivity has been suggested as a possible early-warning signal for the 
onset of instability~\cite{SchefferNature2009,SchefferScience2012,VeraartNature2012,DaiScience2012,DaiNature2013}.
Transient instability not only fills the gap between these two concepts,
but also highlights new consequences of rapid environmental change.
The longer the period of transient instability and the larger the amplification
of perturbations of the population vector, the more susceptible an ecosystem is to multiple perturbations.
One perturbation may drive a stable system into a period of transient instability
that eventually dissipates;
but two or three perturbations in quick succession may force the system
to a new, unknown equilibrium point that may correspond to a loss of species and biodiversity.
Pseudospectra can be used to investigate which ecosystems are at risk of instability,
and what could be done to mitigate that risk.

\section*{Acknowledgments}
We thank Gyuri Barab\'as and one anonymous reviewer for comments
that greatly improved the paper.
FC was supported by Invenia Labs and additionally thanks the London Institute for Mathematical Sciences, and British Ecological Society grant 4785/5824 was awarded to PPAS; 
PPAS was also supported by an AXA Postdoctoral Research Fellowship.
FC and PPAS developed the concept of the study and wrote the paper. FC wrote the software to perform the study.

%\nolinenumbers
%
%\section*{References}
% Either type in your references using
% \begin{thebibliography}{}
% Text
% \end{thebibliography}
%
% OR
%
% Compile your BiBTeX database using our plos2015.bst
% style file and paste the contents of your .bbl file
% here.
% 


\begin{thebibliography}{10}
\bibitem{Afootnote} Eqns~\ref{Eqn:lowerbound}~and~\ref{Eqn:fM} are also valid for bounding normal matrices with positive spectral abscissa. 
As $\epsilon \rightarrow 0$, $\alpha_\epsilon({\bf M})$ converges to the spectral abscissa. 
If ${\bf M}$ has a positive spectral abscissa, then $\lim_{\epsilon\rightarrow 0} \alpha_\epsilon({\bf M})/\epsilon\rightarrow \infty$, 
which confirms that the norm is unbounded and the equilibrium point is unstable.
\end{thebibliography}
\end{document}